\catcode`\@=11 
\def\bigans{b }
\def\revans{r }
\def\litans{l }
\def\answ{r }

\ifx\answ\bigans
\documentstyle[11pt]{article}

\fi
\ifx\answ\revans
\documentstyle[twocolumn,eqsecnum,aps,epsf]{revtex}
\fi
\ifx\answ\litans
 \let\l@r=L
\documentstyle[twocolumn,12pt]{article}
\fi
\newread\epsffilein    
\newif\ifepsffileok    
\newif\ifepsfbbfound   
\newif\ifepsfverbose   
\newdimen\epsfxsize    
\newdimen\epsfysize    
\newdimen\epsftsize    
\newdimen\epsfrsize    
\newdimen\epsftmp      
\newdimen\pspoints     
\def\epsfbox#1{\global\def\epsfllx{72}\global\def\epsflly{72}%
   \global\def\epsfurx{540}\global\def\epsfury{720}%
   \def\lbracket{[}\def\testit{#1}\ifx\testit\lbracket
   \let\next=\epsfgetlitbb\else\let\next=\epsfnormal\fi\next{#1}}%
\def\epsfgetlitbb#1#2 #3 #4 #5]#6{\epsfgrab #2 #3 #4 #5 .\\%
   \epsfsetgraph{#6}}%
\def\epsfnormal#1{\epsfgetbb{#1}\epsfsetgraph{#1}}%
\def\epsfgetbb#1{%
\openin\epsffilein=#1
\ifeof\epsffilein\errmessage{I couldn't open #1, will ignore it}\else
   {\epsffileoktrue \chardef\other=12
    \def\do##1{\catcode`##1=\other}\dospecials \catcode`\ =10
    \loop
       \read\epsffilein to \epsffileline
       \ifeof\epsffilein\epsffileokfalse\else
          \expandafter\epsfaux\epsffileline:. \\%
       \fi
   \ifepsffileok\repeat
   \ifepsfbbfound\else
    \ifepsfverbose\message{No bounding box comment in #1; using defaults}\fi\fi
   }\closein\epsffilein\fi}%
\def\epsfsetgraph#1{%
   \epsfrsize=\epsfury\pspoints
   \advance\epsfrsize by-\epsflly\pspoints
   \epsftsize=\epsfurx\pspoints
   \advance\epsftsize by-\epsfllx\pspoints
   \epsfxsize\epsfsize\epsftsize\epsfrsize
   \ifnum\epsfxsize=0 \ifnum\epsfysize=0
      \epsfxsize=\epsftsize \epsfysize=\epsfrsize
     \else\epsftmp=\epsftsize \divide\epsftmp\epsfrsize
       \epsfxsize=\epsfysize \multiply\epsfxsize\epsftmp
       \multiply\epsftmp\epsfrsize \advance\epsftsize-\epsftmp
       \epsftmp=\epsfysize
       \loop \advance\epsftsize\epsftsize \divide\epsftmp 2
       \ifnum\epsftmp>0
          \ifnum\epsftsize<\epsfrsize\else
             \advance\epsftsize-\epsfrsize \advance\epsfxsize\epsftmp \fi
       \repeat
     \fi
   \else\epsftmp=\epsfrsize \divide\epsftmp\epsftsize
     \epsfysize=\epsfxsize \multiply\epsfysize\epsftmp   
     \multiply\epsftmp\epsftsize \advance\epsfrsize-\epsftmp
     \epsftmp=\epsfxsize
     \loop \advance\epsfrsize\epsfrsize \divide\epsftmp 2
     \ifnum\epsftmp>0
        \ifnum\epsfrsize<\epsftsize\else
           \advance\epsfrsize-\epsftsize \advance\epsfysize\epsftmp \fi
     \repeat     
   \fi
   \ifepsfverbose\message{#1: width=\the\epsfxsize, height=\the\epsfysize}\fi
   \epsftmp=10\epsfxsize \divide\epsftmp\pspoints
   \vbox to\epsfysize{\vfil\hbox to\epsfxsize{%
      \includegraphics{#1}%
      \hfil}}%
\epsfxsize=0pt\epsfysize=0pt}%

{\catcode`\%=12 \global\let\epsfpercent=
\long\def\epsfaux#1#2:#3\\{\ifx#1\epsfpercent
   \def\testit{#2}\ifx\testit\epsfbblit
      \epsfgrab #3 . . . \\%
      \epsffileokfalse
      \global\epsfbbfoundtrue
   \fi\else\ifx#1\par\else\epsffileokfalse\fi\fi}%
\def\epsfgrab #1 #2 #3 #4 #5\\{%
   \global\def\epsfllx{#1}\ifx\epsfllx\empty
      \epsfgrab #2 #3 #4 #5 .\\\else
   \global\def\epsflly{#2}%
   \global\def\epsfurx{#3}\global\def\epsfury{#4}\fi}%
\def\epsfsize#1#2{\epsfxsize}
\let\epsffile=\epsfbox

\begin{document}


\ifx\answ\revans

\else
\begin{flushright} 
April 28, 1997  \\ gr-qc/9704081
\end{flushright}
\renewcommand{\thefootnote}{\fnsymbol{footnote}}
\setcounter{footnote}{1}
\begin{center} 
\fi

\ifx\answ\revans
\title{
A trick for passing degenerate points 
in Ashtekar formulation
 } 
\fi
\ifx\answ\bigans
\baselineskip .35in
\vskip 1.0cm
{\LARGE{\bf 
A trick for passing degenerate points \\
in Ashtekar formulation
}}
\vskip 1.0cm
\baselineskip .25in
{\large
\fi
\ifx\answ\litans
\vskip 0.25cm 
{\Large{\bf
A trick for passing degenerate points \\
in Ashtekar formulation
}}\vskip 0.8cm 
\fi

\ifx\answ\revans
\author{Gen Yoneda 
}
\address{yoneda@mn.waseda.ac.jp\\
Department of Mathematics, Waseda University, 
Okubo 3-4-1, Shinjuku, Tokyo 169, Japan
}
\author{Hisaaki Shinkai}
\address{shinkai@wurel.wustl.edu\\
Department of Physics, Washington University, 
St.Louis, MO 63130-4899, USA
}
\author{Akika Nakamichi}
\address{akika@mn.waseda.ac.jp\\
Gunma Astronomical Observatory,
Ohtomo 1-18-7, Maebashi-shi, Gunma 371, Japan} 
\date{April 28, 1997}
\maketitle
\else
{\sc Gen Yoneda}\footnote{Electronic address: 
yoneda@mn.waseda.ac.jp},
{\sc  Hisaaki Shinkai}\footnote{Electronic address: 
shinkai@wurel.wustl.edu} 
and  
{\sc Akika Nakamichi}\footnote{Electronic address: 
akika@mn.waseda.ac.jp}
\fi

\ifx\answ\bigans
} ~\\ 
\fi
\ifx\answ\litans
\vskip 0.8cm 
\fi
\ifx\answ\revans
\else
$~{\dag}$ {\em  Department of Mathematics, Waseda University, \\
 Okubo 3-4-1, Shinjuku, Tokyo 169, Japan} \\
$~{\ddag}$ {\em  Department of Physics, Washington University, \\
 St.Louis, MO 63130-4899, USA} \\
$~{\mbox \S}$ {\em Gunma Astronomical Observatory,
Ohtomo 1-18-7, Maebashi-shi, Gunma 371, Japan} \\
\fi

\ifx\answ\bigans
\vskip 1.5cm 
\end{center}
\begin{abstract}
\baselineskip .225in
\fi
\ifx\answ\revans
\begin{abstract}
\fi
\ifx\answ\litans
\vskip 0.3cm 
{\bf abstract}
\vskip 0.15cm 
\begin{minipage}[c]{9.5cm}
\small
\fi

We examine one of the advantages of Ashtekar's 
formulation of general relativity: a tractability of degenerate 
points from the point of view of following the dynamics of 
classical spacetime.
Assuming that all dynamical variables are finite, we conclude that an
essential trick for such a  continuous evolution
is in complexifying variables.
In order to restrict 
the complex region locally, we propose 
some `reality recovering' conditions on spacetime.
Using a degenerate solution derived by pull-back technique,
and integrating the dynamical equations numerically,
we show that this idea works in an actual dynamical problem.
We also discuss some features of these applications. 

\ifx\answ\bigans
\end{abstract}
\vfill
\fi
\ifx\answ\revans
\end{abstract}
\fi
\ifx\answ\litans
\end{minipage}
\end{center} 
\vspace{1.0cm }
\fi

\ifx\answ\bigans
\baselineskip = 15pt
\fi
\ifx\answ\revans
\narrowtext
\fi
\ifx\answ\litans
 \vspace{2.0cm }
  \baselineskip = 12pt
\fi
\renewcommand{\thefootnote}{\arabic{footnote}}
\setcounter{footnote}{0}

\section{Introduction} \label{sec:intro}
A decade has passed since the proposal of the new formulation of 
general relativity by Ashtekar \cite{Ashtekar}. 
By using this special pair of variables, the framework has many 
advantages in the treatment of gravity.
The constraint equations which appear in the theory become 
low-order polynomials, and the theory has the
correct form for gauge theoretical features. These 
suggest possibilities for treating a quantum 
description of gravity nonperturbatively.

Here we examine another advantage of the $SO(3)$-ADM 
(Ashtekar) formulation of general relativity: 
the ability to dynamically evolve a spacetime
with a degenerate metric consistently.
A `degenerate point', here,  is defined as a point in the spacetime 
where the density $e$ of 3-space vanishes, as we denote in \S 
\ref{sec:defpass}.
This advantage comes from the fact that all equations in  Ashtekar's 
formulation do not contain the inverses of the variables. 

There are several motivations for studying `degenerate points'. 
The first one comes from a quantum cosmological description of 
the very early history of the Universe.
One scenario describes that our Universe was born and evolved in 
Euclidean spacetime, then emerged with Lorentzian metric through a 
quantum tunneling process. 
Kodama \cite{Kodama} found an exact solution for Bianchi type IX 
spacetime using Ashtekar's framework, which has both Euclidean and 
Lorentzian sections connected by an analytic continuation.
Apparently, such a signature changing process should
pass through a degenerate point, but so far the dynamics
of this process remains
unclear. 

A second motivation comes from the gauge theoretical point of view.
We can set a finite gauge transformation so as to have a degenerate
point of spacetime from a non-degenerate 
one naturally\cite{Horowitz}.
Therefore when we calculate the functional integrals, 
we have to include degenerate solutions due to the 
gauge invariance, and studies on 
degenerate points will be indispensable to construct the 
complete theory of quantum gravity in the near future. 

A third interesting application exists in  
topological field theory. Since 
Ashtekar's dynamical variables $\tilde{E}^i_a$ and ${\cal A}^a_i$
are assigned positive mass dimension one, a short-distance limit 
(i.e., region of quantum gravity) 
must have degenerate phase: $\tilde{E} = {\cal A} = 0$.
Such a degenerate phase is called the topological (unbroken) phase of
quantum gravity,   in which diffeomorphism is unbroken\cite{Witten}.
Finding how to evolve from the degenerate unbroken phase to the 
non-degenerate phase is one of the important issues,
and it is expected to 
be a key  to construct quantum gravity in four dimensions.

Another possibility which makes degenerate points
interesting is the evolution of classical  
(Lorentzian) spacetime. 
As already pointed out by some authors\cite{ys-con,Salisbury},
the Ashtekar formulation is also attractive for 
numerical relativity.  
If this tractability of a degenerate point works also in this
context, then  we will be able to analyze 
focusing or shell-crossing features or (coordinate) singularities 
in spacetime.


In this paper, we study what problems appear and how to solve 
them when we treat the dynamics through a degenerate point.
Here, the  dynamics means the evolution of the three-hypersurfaces 
expressed by Ashtekar's variables ($\tilde{E}^i_a, {\cal A}^a_i$), 
fixing the gauge freedom of the lapse function 
$\null \! \mathop {\vphantom {N}\smash N}\limits ^{}_{^\sim}\!\null$,
shift vector $N^i$ and triad lapse
function ${\cal A}^a_0$ (this name appears in \cite{ys-con}) 
every time step.  
This is a common procedure for 
treating the dynamics in the ADM formulation, though we have 
an additional gauge freedom.
In order to set up a problem of {\it dynamics} 
clearly, we fix our spacetime to have Lorentzian signature.

In the last few years, several studies have been done on  
degenerate metrics.
Some solutions have been found by connecting degenerate and 
non-degenerate metrics
\cite{Bengtsson91}-\cite{Bengtsson93}, and 
degenerate examples in terms of Ashtekar variables 
are shown in \cite{Bengtsson90}. 
Some alternative approaches over the framework of Ashtekar were
considered in \cite{Bengtsson89} and \cite{Jacobson-Romano}.
However there are few comments
on the dynamical evolution of a degenerate metric.
Bombelli and Torrence \cite{Bombelli}  commented on the  
conditions for `passing' degenerate points.
They speculate the degeneracy of lapse function, 
condition for finite `passing' coordinate-time, 
and possibility of divergence of variables at the degenerate point.
We will develop their idea in  \S \ref{subsec:cusp}.  


Assuming that both all dynamical variables and the 
coordinate time are finite ({\it passing condition} in \S 
\ref{sec:defpass}), 
we seek the condition which enables us
to {\it pass} degenerate points  for the case of a vacuum 
spacetime with/without a cosmological constant.
We take mainly two approaches. The first approach, we named
{\it intersecting slice approach}, purposes that a foliation passes 
a degenerate point directly. This is a natural passing behavior
which we expect from a context of Ashtekar's formulation. 
However, we show that 
such a direct treatment does not work normally both in ADM and
Ashtekar formulation, whether we impose the reality conditions 
or not. This  is described in \S \ref{sec:inter}.
The second approach, we call {\it deformed slice approach}, takes 
a complex path around a real degenerate point. 
This approach is described in \S \ref{sec:deform}, 
in which we propose some conditions for recovering the real 
manifold after a detour
around a degenerate point by complexification. 
In \S \ref{sec:example}, we derive a degenerate solution by pull-back
technique and show that the  deformed slice approach works
well for `passing' a degenerate point 
using a numerical integration of the dynamical equations.
We conclude that the essential trick for `passing' a degenerate point
in Ashtekar's formulation exists in their complexification of the
basic dynamical variables.
We devote \S \ref{sec:discuss} to discussion.
  
When we apply the Ashtekar formalism in classical general 
relativity, we need to impose reality conditions. 
We treat these reality conditions based on work developed 
by two of us  \cite{ys-con}, 
and some essential points are attached in the
Appendix along with a full 
description of the notations we use throughout this paper.


We use greek letters ($\mu, \nu, \rho, \cdots$), which range over 
the four spacetime coordinates $0, \cdots, 3$, 
while uppercase latin letters from the middle of
the alphabet ($I, J, K, \cdots$) range over the four internal 
SO(1,3) indices  $(0), \cdots, (3)$.
Lower case latin indices from the middle of the alphabet 
$(i, j, k, ...)$ range over the three spatial indices $1, \cdots, 3$,
while lower case latin indices from the beginning of the alphabet
$(a, b, c, ...)$ range over the three internal SO(3) indices
 $(1), \cdots, (3)$\footnote{We raise and lower 
$\mu,\nu,\rho$ by $g^{\mu\nu}$ and $g_{\mu\nu}$ (Lorentzian metric);
$I,J,K$ by $\eta^{IJ}={\rm diag}(-1,1,1,1)$ and $\eta_{IJ}$;
$i,j,k$ by $\gamma^{ij}$ and $\gamma_{ij}$(3-metric).}.
We use volume forms $\epsilon_{abc}$; 
$\epsilon_{abc} \epsilon^{abc}=3!$. 

\section{Definitions of `passing'}\label{sec:defpass}
In the beginning, let us clarify some terminology.
The basic variables and equations in Ashtekar's formulation are
summarized in the Appendix.

A `degenerate point', in this paper, is the point 
in the spacetime where
the density $e$ of 3-space vanishes (not 4-space).
In the Ashtekar formulation, the density
is defined as $e=\sqrt{\det \tilde{E}^i_a}$, 
which corresponds to the same
condition in the ADM formulation as $e^2=\det \gamma_{ij}$, where
$\gamma_{ij}$ is 3-metric. We ask readers to remember that such a 
degenerate point does not  always mean a physical singularity.
In the basic equations neither the constraints 
(\ref{c-ham})-(\ref{c-g}) nor the
dynamical equations (\ref{eqA})-(\ref{eqE})  include any inverse
of the dynamical variables (${\cal A}^a_i$ and $\tilde{E}^i_a$), 
even if we include a 
cosmological constant 
\footnote{The cosmological constant term in (\ref{eqA}) looks
as if it might be divergent, but  
the relation 
$ e e^a_i = (1/2) \epsilon^{abc}  
\null\!\mathop{\vphantom {\epsilon} \smash \epsilon}
\limits ^{}_{^\sim}\!\null_{ijk}
\tilde{E}^j_b \tilde{E}^k_c $
guarantees its finiteness. }.
This fact suggests to us that we can `pass' such a degenerate point. 
That is, we expect that all calculations can be continued even if we
have a degenerate metric during the time evolution.

Note that, in the ADM formulation, this is impossible because 
the equations in the ADM include  an inverse of the variables. 
We also remark that
we can not transform the Ashtekar variables onto 
the ADM three-hypersurface $\Sigma$ at the degenerate point.
This is because the 3-metric $\gamma_{ij}$
is given by $\gamma_{ij}:=e^a_ie^a_j$, where $e^a_i:=(E^i_a)^{-1}$ 
and $E^i_a:=\tilde{E}^i_a / e$, and the last quantity
diverges as  $e \rightarrow 0$.
Therefore, at the degenerate points, 
one can not recover the ADM variables from Ashtekar variables. 

In order to say `pass' degenerate points, we require the following 
four {\it passing conditions}: \\
\noindent
(a) Ashtekar variables $\tilde{E}^i_a, {\cal A}^a_i, 
\null \! \mathop {\vphantom {N}\smash N}
\limits ^{}_{^\sim}\!\null, N^i, {\cal A}^a_0$
must remain finite throughout the calculation,\\
\noindent
(b) the spatial derivatives of them must also be finite 
throughout the calculation (because they appear in the equations of 
motion),\\
\noindent
(c) all the constraints 
(\ref{c-ham}),(\ref{c-mom}),(\ref{c-g})
and the equations of motion (\ref{eqA}),(\ref{eqE}) 
\footnote{From the assumptions (b) and (c), we have that
the time differentiation of dynamical variables 
$\dot{\tilde{E}^i_a}$, $\dot{{\cal A}^a_i}$ must also be finite. }
are satisfied,
and \\
\noindent
(d) the calculation must be finished in finite coordinate time.

The problem is whether we can make dynamical foliation through a 
degenerate point under these passing conditions
(a)-(d) above.
In the following sections, we take two approaches. 
The first one, which we call the
`intersecting slice approach', attempts to pass 
a degenerate point directly, 
and the second one, `deformed slice approach', 
takes a foliation in complex region.

\section{Intersecting slice approaches}\label{sec:inter}
In this section, we describe the possibility of an 
`intersecting slice approach', which attempts to pass 
a degenerate point directly.
Here `direct' means that the dynamical Ashtekar variables 
run into the degenerate points, never detour the points. 
We assume that $N$ and $N^i$ are real as usual. 
To say our conclusion first, we face 
at least two problems, which we call  ``cusped lapse/density problem"
 and ``divergence problem",  the latter requires severe 
conditions. 

\subsection{Cusped lapse/density problem}\label{subsec:cusp}

A troublesome variable in an evolution through a degenerate point
is the (inverse) densitized lapse 
$\null \! \mathop {\vphantom {N}\smash N}
\limits ^{}_{^\sim}\!\null := N/e_\Sigma$, 
where $e_\Sigma=\sqrt{\det\gamma_{ij}}$.
Since $\null \! \mathop {\vphantom {N}\smash N}
\limits ^{}_{^\sim}\!\null$ is held finite [condition (a)], 
the ADM lapse $N=\null \! \mathop {\vphantom {N}\smash N}
\limits ^{}_{^\sim}\!\null
e_\Sigma$ vanishes at the degenerate point  $e_\Sigma=0$.
Notice that the ratio of the proper time $\tau$ 
to the coordinate time $t$ is 
not $\null \! \mathop {\vphantom {N}\smash N}
\limits ^{}_{^\sim}\!\null$ but $N$.
Thus we are afraid that the calculation exhausts an
infinite amount of 
time, i.e., 
condition (d) is violated.
Let us take $t$ and $\tau$ such that 
$\tau=t=0$ at the degenerate point $e_\Sigma=0$. 
Then condition (d) is denoted by
\footnote{Strictly, this is an improper Riemann integral.}
\begin{equation}
\delta t = \int_{-t_0}^{t_1} dt = 
\int_{-\tau_0}^{\tau_1} {d\tau \over N(\tau)} < \infty.\label{dlpc}
\end{equation}
Here the range of the integral is arbitrary 
but includes the zero point $t=\tau=0$.
Note that $N\geq 0$ since it is a lapse function.
This condition originally appeared in \cite{Bombelli}. 

In terms of $\tau$, the request (\ref{dlpc}) 
is considerably restrictive
because $N=0$ at $\tau=0$.
For example, let us consider the form
$N=|\tau|^s$ where $s$ is constant.
The condition for (\ref{dlpc}) is $0<s<1$,
e.g., $N=\sqrt{|\tau|}$ satisfies (\ref{dlpc}).
We see $dN/d\tau$ does not exist at $\tau=0$
\footnote{
We see
$\limsup_{\tau\to 0+} dN/d\tau =\infty $
when $d N / d\tau$ exists for $0<\tau$.}.
This is the reason we call this the {\it cusped lapse problem}.
We note that this is not  a serious problem
to execute, because the choice of lapse in such a cusped
form is only within a freedom of gauge, although there
remains its naturalness of foliation.

In terms of $t$, however, the request (\ref{dlpc}) 
is not restrictive as follows.
When $N=d\tau/dt=\tau^s$ $(0<s<1)$, 
we see $t=\tau^{1-s}/(1-s)$,
thus 
$N=(-s+1)^{s/(1-s)}t^{s/(1-s)}$.
The power $s/(1-s)$ can be taken arbitrary positive
even if we require $0<s<1$.
The only request here is 
$N=0$ at $t=0$.
So the lapse is not necessarily cusped in terms of $t$.

Furthermore we will see that
the density $|e|$ should also satisfy a
condition which is similar to  the
cusped lapse condition. 
We call this problem the {\it cusped density problem}.
We assume $e=0$ at $t=\tau=0$ and $|\null 
\! \mathop {\vphantom {N}\smash N}\limits ^{}_{^\sim}\!\null|<M$ 
(bounded).
Since $N\geq 0$, we have 
$e \null \! \mathop {\vphantom {N}\smash N}
\limits ^{}_{^\sim}\!\null=|e \null \! 
\mathop {\vphantom {N}\smash N}\limits ^{}_{^\sim}\!\null|
=|e| \ | \null \! \mathop {\vphantom {N}\smash N}
\limits ^{}_{^\sim}\!\null|$.
Then we see
\[
\infty 
> 
\int^{\tau_1}_{-\tau_0} \displaystyle{d\tau\over N}
=
\int^{\tau_1}_{-\tau_0} \displaystyle{d\tau \over e 
\null \! \mathop {\vphantom {N}\smash N}\limits ^{}_{^\sim}\!\null}
=
\int^{\tau_1}_{-\tau_0} \displaystyle{d\tau \over |e| 
\  |\null \! \mathop {\vphantom {N}\smash N}
\limits ^{}_{^\sim}\!\null|}
>
\displaystyle{1\over M}
\int^{\tau_1}_{-\tau_0} \displaystyle{d\tau \over |e|}.
\]
Similarly to the lapse, the density is 
restricted in terms of $\tau$
as $|e|=\tau^{s}$ $(0<s<1)$,
and is not restricted in terms of $t$.

\subsection{Divergence problem}\label{subsec:div}
Let us consider the quantity $\omega^{0a}_i$
which appears in the definition of ${\cal A}^a_i$ 
[(\ref{w2A}) in Appendix].
We have
\[
\omega^{0a}_i=-K_{ij}E^{ja}
=\displaystyle{1\over e^2 \null \! \mathop {\vphantom {N}\smash N}
\limits ^{}_{^\sim}\!\null}(\dot{\gamma_{ij}}-D_iN_j-D_jN_i)
\tilde{E}^j_a.
\]
In order to examine this finiteness, we prepare 
\begin{eqnarray}
\omega^{0a}_i\tilde{E}^i_a
&=&
\displaystyle{1\over  \null \! \mathop {\vphantom {N}\smash N}
\limits ^{}_{^\sim}\!\null}\left(
\displaystyle{\partial_t(e^2)\over e^2}
-D_i(N^i) -D_j(N^j)
\right) \nonumber \\ &=&
\displaystyle{1\over  \null \! \mathop {\vphantom {N}\smash N}
\limits ^{}_{^\sim}\!\null}\left[
\displaystyle{\partial_t(e^2)\over e^2}
-\displaystyle{\partial_j(e^2)\over e^2}N^j
-2\partial_i(N^i)
\right]. \label{prepare}
\end{eqnarray}
Now we assume the conditions (a)-(d).
Then $\null \! \mathop {\vphantom {N}\smash N}
\limits ^{}_{^\sim}\!\null$ is finite,
so the term $1/\null \! \mathop {\vphantom {N}\smash N}
\limits ^{}_{^\sim}\!\null$ is bounded below.
When a parameter $s$ is taken such that
$\partial_s=\{1,-N^i\}$ and $s=0$ corresponds the degenerate point,
the first and second terms in the right-hand side  
of (\ref{prepare}) are rewritten as, 
\[
\displaystyle{\partial_t(e^2)\over e^2}-
\displaystyle{\partial_j(e^2)\over e^2}N^j
={\partial_s (e^2) \over e^2}
= \partial_s {\log (e^2)},
\]
which diverges at $s=0$. 
The $\partial_i(N^i)$ term in (\ref{prepare})
is finite since we assume condition (b).
To sum up, 
we see (\ref{prepare}) diverges at the degenerate point. 
If we assume the triad reality conditions 
(\ref{s-reality1})-(\ref{s-reality2}), 
this fact tells us the passing is impossible.
Here the triad reality conditions are useful and stronger than
the usual metric reality conditions as in Appendix A.2.
Since the triad is real, $\omega^{0a}_i$ is a real part of 
${\cal A}^a_i$,
so it is finite.
And, since $\tilde{E}^i_a$ is finite, $\omega^{0a}_i\tilde{E}^i_a$
must be finite.
The above conclusion contradicts this.
Thus, unfortunately,
we can not pass the degenerate point
when the triad reality is assumed.

Even if we do not impose the triad reality conditions, i.e.
we impose the metric reality conditions 
or do not impose any reality conditions, 
this problem still exists as follows.
In order to enforce the finiteness of ${\cal A}^a_i$,
its second term, $i{\epsilon^a}_{bc}\omega^{bc}_i$, 
must diverge such that
it cancels out the divergence of $\omega^{0a}_i$.

At the present stage, we do not know any exact solutions
which satisfy this condition. 
Thus we conclude that 
the passing conditions (a)-(d), which aimed to pass a degenerate
point directly, are difficult to satisfy simultaneously
 due to this divergence of variable.

\section{Deformed slice approaches} \label{sec:deform}

In the previous section, we showed that foliations through 
a degenerate point are difficult even within the Ashtekar framework.
Therefore, 
in the second approach, we try to make a spacetime foliation avoid a
degenerate point by breaking reality conditions.
That is, we impose the foliations of 3-space which detour 
into complex spacetime only 
in the vicinity of such a degenerate point. 
This means that locally we are out of Einstein's framework. 
In this section, we examine what conditions are
needed
when we impose
such a locally deforming time evolution. 

As we mentioned in the introduction, our aim is to keep
continuous time 
evolution in the presence of a degenerate point. 
We extend 
the
spacetime using complex numbers,
and impose reality 
on the coordinates
$(t,\mbox{\boldmath$x$})$,
so the foliation maintains 
$3 + 1$ dimensions 
$\mbox{\boldmath $R$}^3 \times \mbox{\boldmath $R$}$ in 
$\mbox{\boldmath $C$}^4$.
Note that our proposal of real coordinate 
 is different from the Wick rotation, rather it is
close to the recent proposal of complex lapse function by Hayward 
\cite{Hayward}. We use a freedom of gauge function to foliate into 
the complex spacetime. 

We assume a spacetime has a degenerate point and has 
also its future (a kind of coordinate singularity or focusing).
In the following expression, we postulate that a 
single degenerate point 
exists at $(t,\mbox{\boldmath$x$})=(t_*,\mbox{\boldmath$x$}_*)$, but 
extending to many degenerate points is straightforward. 
In order to recover the Einstein spacetime, the foliations have
to satisfy the following conditions, such as 
the `foliation recovering' 
condition which ensures
that the foliation locally deforms into the imaginary region only  
in the vicinity of the degenerate point. 

\subsection{Foliation recovering condition} \label{subsec:fr}
We seek the foliation which 
starts from the real section, evolves in the complex region 
only in the vicinity of a degenerate point, and ends in the 
real section. 
We call this the `foliation recovering'.
In the local deformation, 
the lapse function and the shift vector become complex
valued 
to foliate in the complex region. 
Thus the `foliation recovering condition' is expressed 
in terms of the gauge functions as:  

\begin{eqnarray}
\int^{t_+}_{t_-} \Im N(t,\mbox{\boldmath$x$}) dt&=&0, 
\label{recoverC} \\
\int^{t_+}_{t_-} \Im N^i(t,\mbox{\boldmath$x$}) dt&=&0, 
\label{recovershift}
\end{eqnarray}
where $t_-<t_*$ and $t_+>t_*$ are defined in appropriate far time
from a degenerate point at $t=t_*$.
The reason for not imposing the other gauge function, the triad 
lapse ${\cal A}^a_0$, is that the actual spacetime foliation
is 
determined only by the lapse and the shift vector, and the triad 
lapse does not contribute in physical (external spacetime) 
foliations. 

\subsection{Asymptotic reality condition}\label{subsec:asympt}

Although we are considering deforming Einstein spacetime by using
a complex  metric in the time evolution, we require 
that such deformations
are as local as possible. This is because our
proposal is just a technique for treating a degenerate point.
Therefore we impose reality conditions both at 
the spatial far limit
and time far limit from a degenerate point, 
 in order to ensure that the 
 metric becomes complex only in the vicinity of the degenerate
 points. 

In the spatial far region from a degenerate
point, the spacetime must be in real Einstein spacetime. 
In the time far limit, we also impose a condition 
to ensure that the metric becomes complex 
only in the time vicinity of the degenerate point. 
In other words, 
the initial value of $\tilde{E}^i_a (t,\mbox{\boldmath$x$})$
is chosen to be real.
Then we make time evolution within Einstein spacetime before we meet
a degenerate point, and in the vicinity of a degenerate point
we make time evolution into complex values, 
and return to the real values at the end.

Those are expressed by gauge functions and metric variables. 
The former are:
\begin{eqnarray}
&&\Im N(t,\mbox{\boldmath$x$}) \rightarrow 0 \label{asymp-gauge1}\\
&&\Im N^i(t,\mbox{\boldmath$x$}) \rightarrow 0 \label{asymp-gauge2}
\end{eqnarray}
for all four limits 
$\mbox{\boldmath$x$} \rightarrow \mbox{\boldmath$x$}_* 
\pm \Delta \mbox{\boldmath$x$}$,
$t \rightarrow t_* \pm \Delta t$
where $\Delta \mbox{\boldmath$x$}$ and $\Delta t$
determine 
where we call the
{\it far region} from a degenerate point, which depends on 
the particular problem. The latter is: 
\begin{equation}
\Im \left[ {\tilde{E}^i_a}{\tilde{E}^{j a}} / {\rm det} \tilde{E} 
(t,\mbox{\boldmath$x$}) \right]  \rightarrow 0
\label{asymp-metric} 
\end{equation}
for all four limits 
$\mbox{\boldmath$x$} \rightarrow \mbox{\boldmath$x$}_* 
\pm \Delta \mbox{\boldmath$x$}$,
$t \rightarrow t_* \pm \Delta t$.
This expresses the primary reality condition
(\ref{w-reality1}), of which general discussions are in Appendix. 

As for the time-far limit of the metric reality condition 
(\ref{asymp-metric}),
the expression can be written in an integrated form 
\begin{equation}
\int^{t_+}_{t_-} \frac{d}{dt}  
\left[ \Im{\tilde{E}^i_a}{\tilde{E}^{j a}} 
 / {\rm det} \tilde{E} (t,\mbox{\boldmath$x$})dt \right] =0. 
\label{Tasymp}
\end{equation}
The integrand is calculated using the dynamical evolution equation
(\ref{eqE}), so the explicit form of (\ref{Tasymp}) is not given
at an arbitrary time except in some simple analytic cases
\footnote{Of course the  reality conditions at an arbitrary time 
include this asymptotic condition. 
But reality conditions at an arbitrary time 
are too restrictive to solve in the deformed slice approach.}.
Normally, the condition (\ref{Tasymp}) becomes a boundary problem of
the evolution both in initial and final time-slice, 
so that 
any general application of
this method requires numerical iterations for 
finding gauge functions to satisfy all conditions 
(\ref{recoverC})-(\ref{asymp-metric}).

In general,
we cannot deny the possibility that two
different gauges satisfying (\ref{recoverC})-(\ref{asymp-metric})
exist and each yields a different real metric after deforming.
However, there is no problem when we know the metric 
beyond the degenerate point.
In the next section, we show an example of such a foliation, where
we fix our background metric to be flat spacetime with one 
degenerate point.

\section{Examples}\label{sec:example}

In this section, we discuss dynamical evolution through a 
degenerate point, using an analytic solution with the
pull-back 
technique and a numerical demonstration.


\subsection{Exact solution with degenerate point}\label{subsec:exact}

One method to transform an exact solution is 
by the pull-back technique.
When pull-back includes a singularity, it is not a diffeomorphism, 
and
the coordinate transformation from the non-degenerate metric
to the degenerate metric is singular. 
This technique was also used by 
Horowitz \cite{Horowitz}, when he discussed a possibility of 
topology change of the spacetime. 
Here, we derive a flat Minkowski spacetime which has one degenerate
point at the origin by using the pull-back method for later 
numerical demonstration of deforming slices. From the
flat 
Minkowski metric, the following degenerate metric is 
constructed:
\begin{eqnarray}
ds^2&=&
-[1-(f'(t)h(x))^2]dt^2
\nonumber \\ &&
+[-2f'(t)h(x) (1-f(t)h'(x))]dtdx
\nonumber \\ &&
+(1-f(t)h'(x))^2dx^2+dy^2+dz^2,\label{degmet}
\end{eqnarray}
where $f(t),h(x)$ are an arbitrary functions which
satisfy
\begin{eqnarray}
&&f(\pm\infty)=f'(\pm\infty)=0, \quad  -1<f'(t)<1, \nonumber \\
&&\mbox{and~~}-1< f(t) \leq 1, 
\label{condf}
\end{eqnarray}
where the  
equality in the last equation is satisfied only when $t=0$, and
\begin{eqnarray}
&&h(\pm\infty)=h'(\pm\infty)=0, \quad -1<h(x)<1, \nonumber \\
&&\mbox{and~~}-1<h'(x)\leq 1, 
\label{condh}
\end{eqnarray}
where the  
equality in the last equation is satisfied only when $x=0$.

We will see that this metric has the following properties: 
(1$^\circ$) flat, 
(2$^\circ$) asymptotic Minkowski, 
(3$^\circ$) degenerate at $t=x=0$, and 
(4$^\circ$) Lorentzian ($t$-time, $x,y,z$-space) at elsewhere.
\\
{\it Proof)}\\
(1$^\circ$) Let us put 
$a=t$, $b=x-f(t)h(x)$.
Then we see
$ds^2=-da^2+db^2+dy^2+dz^2$. \\
(2$^\circ$)
Since 
$f(t)\to 0$ and 
$f'(t)\to 0$ when $t\to\pm\infty$,
$h(x)\to 0$ and 
$h'(x)\to 0$ when $x\to\pm\infty$,  
we see that the metric becomes 
flat for all four limits $t\to\pm\infty$, $x\to\pm\infty$.
\\
(3$^\circ$)
The determinant of this metric is
$\det g_{\mu\nu}=-(1-f(t)h'(x))^2$.
Noting 
$-1< f(t) \leq 1$ (equality is satisfied only when $t=0$) and 
$-1< h'(x) \leq 1$ (equality is satisfied only when $x=0$),
we obtain 
$\det g_{\mu\nu} \leq 0$
(equality is satisfied only when $t=x=0$).
\\
(4$^\circ$)
We have
$g_{00}=-1+(f'(t)h(x))^2$.
Since $|f'(t)|<1$, $|h(x)|<1$,
we obtain
$g_{00}<0$.
We easily see $g_{11}\geq 0$.
And we have
$\det g_{\mu\nu}\leq 0$. 
Thus we see the metric is 
Lorentzian ($t$-time, $x,y,z$-space) except for $t=x=0$.

For example, 
$f(t)=e^{-t^2}$, $h(x)=x e^{-x^2}$
satisfy conditions (\ref{condf}) and (\ref{condh}).
We start from the metric given by the substitution of 
$f(t)=e^{-t^2}$, $h(x)=x e^{-x^2}$
into (\ref{degmet}), 
\begin{eqnarray}
ds^2&=&
-[1-(2txe^{-t^2-x^2})^2]dt^2
\nonumber \\ &&
+4txe^{-t^2-x^2}[1-(1-2x^2)e^{-t^2-x^2} ]dtdx
\nonumber \\&&
+[1-(1-2x^2)e^{-t^2-x^2}]^2 dx^2+dy^2+dz^2. \label{dM}
\end{eqnarray}
Hereafter, we call this the dM solution 
(a degenerate expression of Minkowski spacetime).
This metric is plane symmetric so that a degenerate point
$t=x=0$ is not a point in spacetime.  
But this metric will be sufficient to show our proposal
works in the next demonstration. 

\subsection{Numerical demonstration of two approaches}
\label{subsec:numerical}

Using the dM solution (\ref{dM}) as a background metric, 
we here show some numerical demonstration of our discussions in the
previous sections. 
The following computations were made by conventional 
finite difference
scheme using independent programs written in Fortran and 
Mathematica,
both of which passed the
appropriate convergence tests of grid size 
using the final imaginary part of
density $e$.

Since the metric (\ref{dM}) has a degenerate point at $x=t=0$,
it is clear that we can not evolve
the solution beyond a degenerate point 
because the inverse 3-metric 
diverges at the degenerate point within the ADM formulation.
As we denoted in \S \ref{sec:inter}, even if we take 
the Ashtekar formulation in time evolutions, 
some variables diverge at a degenerate point 
because of the `divergence problem'.
Thus we may say it is hard to `pass' such a point. 

Fig. \ref{ash-inter} shows such a feature. 
The figure is given by a time evolution of 
a solution (\ref{dM}) using Ashtekar's dynamical equation,
and the real part of density $e$ 
is plotted versus spatial and time
comoving coordinates $x$ and $t$. 
We set the gauge conditions (slicing conditions) the
same as the exact solution  (\ref{dM}), i.e., the lapse $N=1$, 
the shift vector $N_i=2txe^{-t^2-x^2}[1-(1-2x^2)e^{-t^2-x^2} ]$, and
the triad lapse ${\cal A}^a_0=0$. 
Initial data are constructed from the exact solution (\ref{dM})
taking triad $e^a_i$ as the diagonal matrix  
$(\sqrt{\gamma_{xx}},1,1)$.
We see that the time evolutions do not work properly
after intersecting $t=x=0$,
because of the `divergence problem'.

\ifx\answ\revans
\begin{figure}[h]
\setlength{\unitlength}{1in}
\begin{picture}(3.3,2.8)
\put(0.3,-0.3){\epsfxsize=3.0in \epsffile{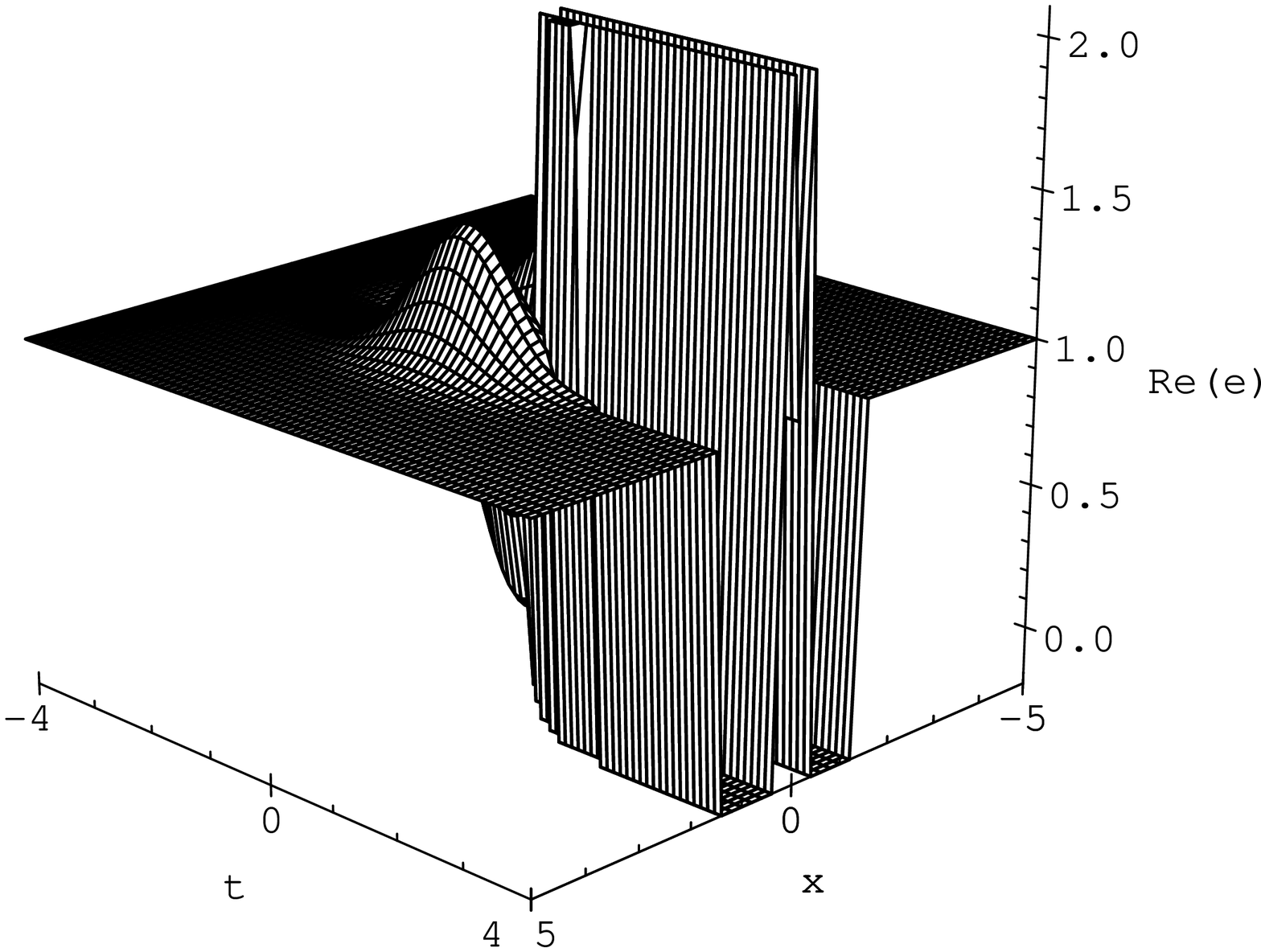} }
\end{picture}
\caption[ash-inter]{
An example of the {\it intersecting approach} for the dM metric 
(\ref{dM})
using Ashtekar's dynamical equations in evolutions. 
The density
is plotted versus spatial and time
comoving coordinates $x$ and $t$. 
A degenerate point exists at the origin
($t=x=0$), and we see that the time evolutions do not work properly
after intersecting the point.
}
\label{ash-inter}
\end{figure}
\fi

Next, we show that a `deformed slicing' approach works well. 
In order to make a deformed foliation, 
we allow  the shift vector 
to be complex. 
The imaginary parts of  $N_i$ are arbitrary but should satisfy 
conditions (\ref{recovershift}) and (\ref{asymp-gauge2}).
As an example, we choose 
\begin{equation}
\Im N_i = a t e^{-b(t^2+x^2)}, \label{im-shift1} \\ 
\end{equation}
where $a$ and $b$ are arbitrary constants.
We take the lapse $N=1$ and the
triad lapse ${\cal A}^a_0=0$, the simplest ones as before.

\ifx\answ\revans
\begin{figure}[b]
\setlength{\unitlength}{1in}
\begin{picture}(3.3,5.4)
\put(0.3,2.6){\epsfxsize=3.0in \epsffile{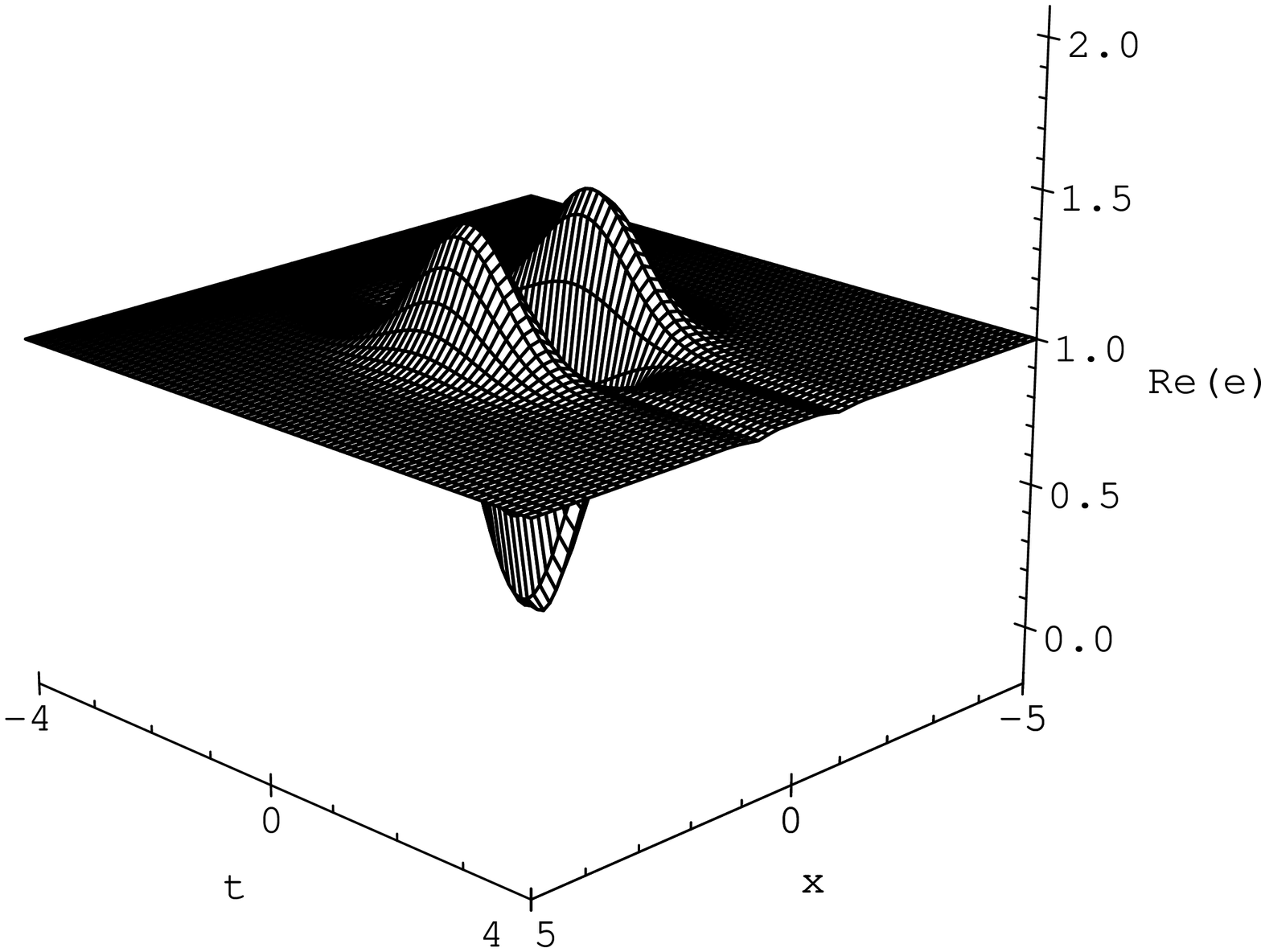} }
\put(0.3,-0.3){\epsfxsize=3.0in \epsffile{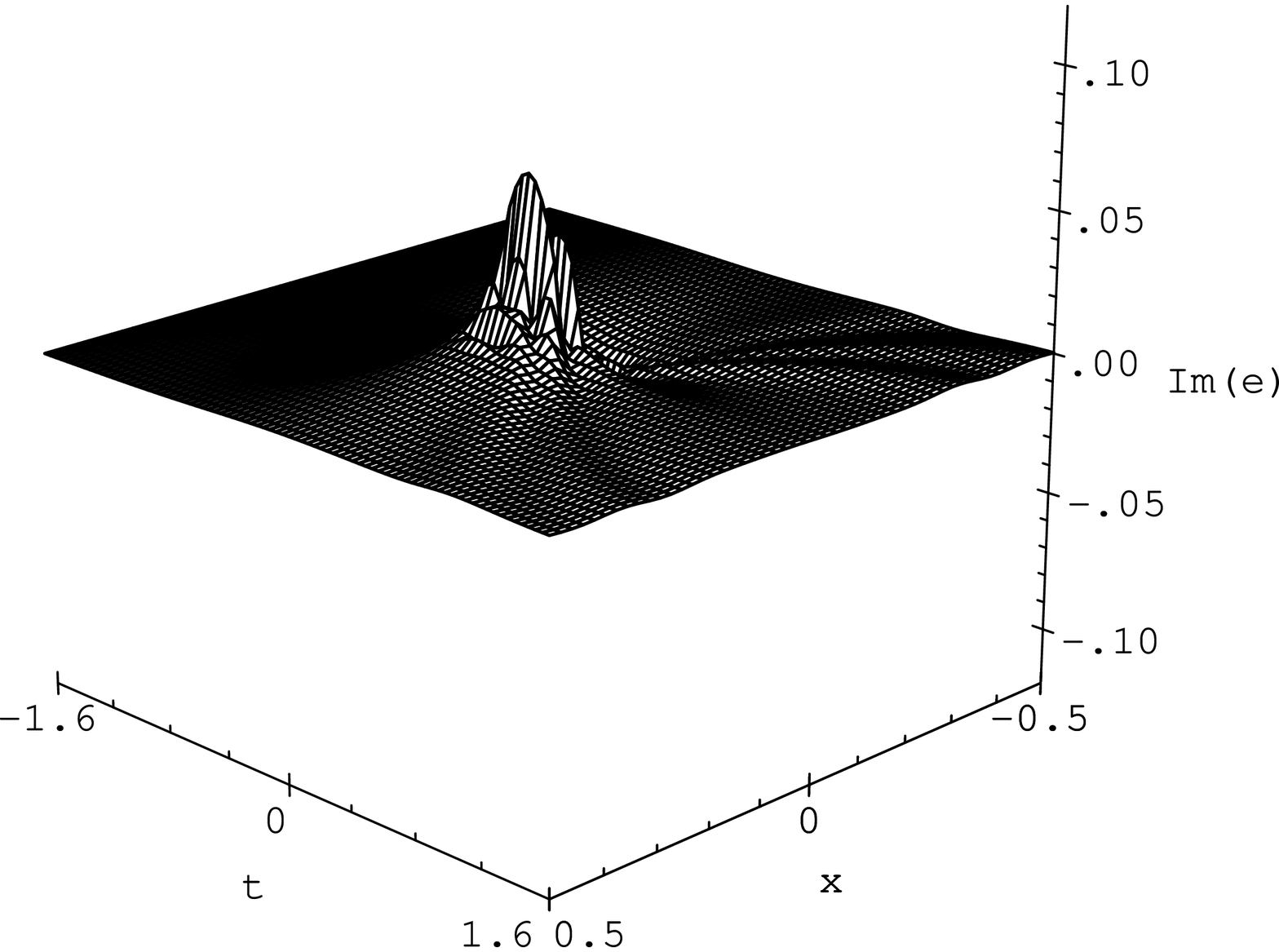} }
\end{picture}
\caption[ash-defo]{
An example of the {\it deformed slicing approach} for the 
dM metric (\ref{dM})
using Ashtekar's dynamical equation in evolution. 
The density
[real part(a) and imaginary part(b)] 
is plotted versus coordinates $x$ and $t$. 
A degenerate point exists at the origin
($t=x=0$), and we see that the time evolution works properly
for all the evolution and that the 3-metric satisfies
 the asymptotic reality condition again. 
}
\label{ash-defo}
\end{figure}
\fi

In order to judge whether an evolution 
recovers asymptotic flatness and 
satisfies asymptotic reality conditions, 
we checked two 
values at the final slice of our evolution; 
\begin{eqnarray}
F(t_{final})&:=& \max_{x} | \Re \left( e (t=t_{final},x) -1 
\right) |
\\
R(t_{final})&:=& \max_{x} | \Im  \left( e (t=t_{final},x)
\right) 
|
\end{eqnarray}
where 
these indicate the time asymptotic reality condition 
(\ref{asymp-metric}). 
($F$ and $R$ were used in convergence tests of the programs.)

By changing a deformation of slice (two parameters $a$ and $b$),
we searched for a solution which minimize the flatness $F$ and 
asymptotical reality $R$ at the final time slice.  
Fig.\ref{ash-defo} shows a successful example of a local
violation of reality, in which we used 
$a=0.003982$ and $b=2.5$ 
in the imaginary part of shift vector (\ref{im-shift1}), and
set $t_{initial}=-4, t_{final}=+4$ in coordinate time. In the case
of Fig.\ref{ash-defo}, the flatness $F$ and asymptotic reality $R$ 
are
$F<O(10^{-2}), R<O(10^{-2})$ respectively when we take 
$\Delta t=0.01$. 
We also should mention that
this example satisfies asymptotic reality conditions at 
spatial far limit and foliation recovering conditions
in the previous section.
In the figure, density $e$
is plotted versus coordinates $x$ and $t$. 
We see that the time evolution does work properly 
in the sense that
the real part recovers the analytic evolutions 
and 
the imaginary part vanished asymptotically.

We found such a solution appears discretely in 
the two-parameter $a$-$b$
plane. This eigensystem-like behavior is what we expected, 
since our requirements are over the freedom
of this dynamical system as we discussed in \S \ref{sec:deform}.
This suggests that the spacetime dynamics in the complexified
manifold will not necessary develop (or converge) 
into the real section of the 
manifold, without imposing some reality recovering conditions. 
And note that a solution which 
satisfies asymptotic reality also satisfies 
asymptotic flatness.

We also tried with other choices of shift vector, e.g.
$\Im N_i = a t x e^{-b(t^2+x^2)}$, 
$\Im N_i = 2 a tx e^{-(t^2+x^2)} 
                \{ 1 - 2 ( 1 - 2 x^2 )  e^{-b(t^2+x^2)}  \} $ 
and other forms
where $a$ and $b$ are arbitrary constants. 
We found that the general behavior of the discrete appearance of 
solutions 
is quite similar to the above example. 
The existence of solutions in the case of small $b$ suggests 
to us that 
our deformed slicing approach is applicable not 
only for degenerate `point' but also for degenerate `region'.

\section{Discussion}\label{sec:discuss}

We have studied one of the advantages of Ashtekar's 
formulation: a tractability of degenerate points.
We showed that a direct passing of degenerate point 
 (which we call the
 intersecting slice approach) is hard to work in general 
 if we impose some natural assumptions 
on dynamical 
variables to express Lorentzian spacetime evolution. 
This means we can not make an evolution of hypersurface directly 
through a degenerate point even if we use Ashtekar's
variables.

Therefore we propose a deformed slice approach, which 
violates reality conditions locally near a 
 degenerate point, 
 and showed that 
this proposal  works in an actual dynamical problem. 
Therefore we conclude that a trick for dynamically passing 
a degenerate 
point exists in using {\it complexified} spacetime.
This deformed slice approach 
is a natural foliation within Ashtekar's formulation since 
 variables are originally defined as 
 complex variables.

Readers may think that if the essential trick of passing 
a degenerate point exists in complexifying variables, then
we might find an example even within the complexified  ADM 
formulation. 
We found also that this statement is true, 
and found similar discrete behavior in the deformed parameters.

This result suggests to us that there is a new possibility for 
dynamical problems in classical spacetime such as 
focusing or shell-crossing of coordinate points and also for
singularities.
Although this foliation requires parameter tuning to satisfy
reality recovering conditions, the fact that a real part of the 
solution always expresses the expected analytic solution indicates
this dynamical evolution technique works for more general problems.

There are some points which remain unclear
 when we break reality conditions locally. 
One big problem is a causality of this spacetime. 
Matschull\cite{Mat} commented on causality 
after a system passes a degenerate point.
He classified the degenerate tetrad by rank, and 
considers causal structure of each degenerate space-time
in a real manifold. 
However  we do not know whether such a usual causal structure can be 
extended to the complex manifold.


If general relativity is extended to allow degenerate
metrics,
the topology of spacetime has a possibility to change even within
a classical picture \cite{Horowitz}. If we allow  breakings of 
reality conditions locally, 
then our classical  path may connect the spaces with different
topology dynamically.
We showed only the case that a spacetime will recover 
its original metric, but in general initial and final metrics 
are allowed to be different.
Further analysis will broaden 
our research in dynamics which includes a
degenerate point. 

A discrete appearance of solutions suggests the existence of 
something like `topological charge' of this system.  We are now
seeking such a charge, and plan to connect our discussions with 
a classically topology changing scenario and/or signature changing
scenario in the early history of the Universe.

We believe that this work is the first example to display 
some classical dynamical behavior of Ashtekar's
variables for 
numerical evolution of spacetime.   
We expect our approach will become the first step 
towards understanding
dynamics of the signature changing process, topology changing
process and causal structure in a complex manifold.


\vspace{0.4cm}

\section*{Acknowledgments}
We thank Hitoshi Ikemori for useful discussions. 
We thank Paul Haines 
for his careful reading of our manuscript.
This work was supported partially by the Grant-in-Aid for Scientific
Research Fund of the Ministry of Education, Science, Sports and 
Culture No. 07854014, by NSF 96PHYS-00507, 
by the Grant-in-Aid for JSPS Fellow (AN) 
and by a Waseda University Grant for 
Special Research Projects 96A-153 and 96A-280.

\appendix
\section{Ashtekar's formulation}
Here, we summarize the basic variables and equations in Ashtekar's
formulation of general relativity, used in this paper. 

\subsection{The Ashtekar variables}\label{subsec:ashva}

The key feature of  Ashtekar's formulation of general relativity
\cite{Ashtekar} is the introduction of a self-dual 
connection as one of the basic dynamical variables.
Let us write the metric $g_{\mu\nu}$ using the tetrad, $e^I_\mu$, 
and define its inverse, $E^\mu_I$, by 
$g_{\mu\nu}=e^I_\mu e^J_\nu \eta_{IJ}$ and $
E^\mu_I:=e^J_\nu g^{\mu\nu}\eta_{IJ}$.
We define a $SO(3,C)$ self-dual connection
\begin{equation} ~^{\pm\!}{\cal A}^a_{\mu} 
:= \omega^{0a}_\mu \mp \displaystyle{i \over 2} 
\epsilon^a_{~bc}\omega^{bc}_\mu,  
\label{w2A}
\end{equation}
where $\omega^{IJ}_{\mu}$ is a spin connection 1-form (Ricci 
connection), $\omega^{IJ}_{\mu}:=E^{I\nu} \nabla_\mu e^J_\nu.$
Ashtekar's plan is to use  only a self-dual part of
the connection 
$^{+\!}{\cal A}^a_\mu$ 
and to use its spatial part $^{+\!}{\cal A}^a_i$ 
as a dynamical variable. 
Hereafter, 
we simply denote $^{+\!}{\cal A}^a_\mu$ as ${\cal A}^a_\mu$.

Note that the extrinsic curvature, 
$K_{ij}=-(\delta_i^{~l}+n_in^l)\nabla_ln_j$
in the ADM formalism, where $\nabla$ is a covariant 
derivative on $\Sigma$, 
satisfies the relation $-K_{ij}E^{ja}=\omega^{0a}_{i}$, 
when the gauge condition $E^0_a=0$ is fixed.
So ${\cal A}^a_{i}$ is also expressed by
\begin{equation} {\cal A}^a_i = -K_{ij}E^{ja}
-\displaystyle{i \over 2} \epsilon^a_{~bc}\omega^{bc}_i. 
\label{def-ash}
\end{equation}

The lapse function, $N$, and shift vector, $N^i$,
are expressed as $E^\mu_0=(1/N, -N^i/N$). 
This allows us to think of
$E^\mu_0$ as a normal vector field to $\Sigma$ 
spanned by the condition $t=x^0=$const.,  
which plays the same role as that of ADM.
Ashtekar  treated the set  (${\cal A}^a_{i}$, $\tilde{E}^i_{a}$) 
as basic dynamical variables, where 
$\tilde{E}^i_{a}$ is an inverse of the densitized triad 
defined by 
\begin{equation} \tilde{E}^i_{a}:=e E^i_{a}, \end{equation}
where $e:=\det e^a_i$ is a density.
This pair forms the canonical set.

In the case of pure gravitational spacetime, 
the Hilbert action takes the form 
\begin{eqnarray}
S&=&\int {\rm d}^4 x 
[ \dot{{\cal A}}^a_{i} \tilde{E}^i_{a}
+\displaystyle{i\over 2} \null \! \mathop {\vphantom {N}\smash N}
\limits ^{}_{^\sim}\!\null \tilde{E}^i_a 
\tilde{E}^j_b F_{ij}^{c} \epsilon^{ab}_{~~c} 
- \Lambda \null \! \mathop {\vphantom {N}\smash N}
\limits ^{}_{^\sim}\!\null \det\tilde{E} \nonumber \\ &&
-N^i F^a_{ij} \tilde{E}^j_a  
+{\cal A}^a_{0}{\cal D}_i \tilde{E}^i_{a} ],
 \label{action}
\end{eqnarray}
where $\null \! \mathop {\vphantom {N}\smash N}
\limits ^{}_{^\sim}\!\null := e^{-1}N$, $\Lambda$ 
is the cosmological constant
\footnote{We changed a factor in front of cosmological
constant from \cite{ys-con}.},  
${\cal D}_i \tilde{E}^i_{a}
    :=\partial_i \tilde{E}^i_{a}
-i \epsilon_{ab}^{~~c}{\cal A}^b_{i}\tilde{E}^i_{c}$, and 
${\rm det}\tilde{E}$ is defined to be  
${\rm det}\tilde{E}=\displaystyle{1\over 6}
\epsilon^{abc}
\null\!\mathop{\vphantom {\epsilon}\smash \epsilon}
\limits ^{}_{^\sim}\!\null_{ijk}\tilde{E}^i_a \tilde{E}^j_b 
\tilde{E}^k_c$, where
$\epsilon_{ijk}:=\epsilon_{abc}e^a_ie^b_je^c_k$
 and $\null\!\mathop{\vphantom {\epsilon}\smash \epsilon}
\limits ^{}_{^\sim}\!\null_{ijk}:=e^{-1}\epsilon_{ijk}$
\footnote{$\epsilon_{xyz}=e$, 
$\null\!\mathop{\vphantom {\epsilon}\smash \epsilon}
\limits ^{}_{^\sim}\!\null_{xyz}=1$, 
$\epsilon^{xyz}=e^{-1}$, 
$\tilde{\epsilon}^{xyz}=1$. }.

Varying the action with respect to the non-dynamical variables 
$\null \! 
\mathop {\vphantom {N}\smash N}\limits ^{}_{^\sim}\!\null$, 
$N^i$ 
and ${\cal A}^a_{0}$ yields the constraint equations, 
\begin{eqnarray}
{\cal C}_{H} &=& 
 \displaystyle{i\over 2} \epsilon^{ab}_{~~c} 
\tilde{E}^i_{a} \tilde{E}^j_{b} F_{ij}^{c} 
  -\Lambda \det\tilde{E}
   \approx 0, \label{c-ham} \\
{\cal C}_{M i} &=& 
  -F^a_{ij} \tilde{E}^j_{a} \approx 0, \label{c-mom}\\
{\cal C}_{Ga} &=&  {\cal D}_i \tilde{E}^i_{a}  
 \approx 0,  \label{c-g}
\end{eqnarray}
where ${F}^a_{\mu\nu} := (d {\cal A}^a)_{\mu\nu}
-\displaystyle{i\over 2}{\epsilon^a}_{bc}({\cal A}^b 
 \wedge {\cal A}^c)_{\mu\nu}$
is the curvature 2-form. 

The equations of motion for the dynamical variables
(${\cal A}^a_i$ and $\tilde{E}^i_a$) are 
\begin{eqnarray} \dot{{\cal A}}^a_{i} &=&
-i \epsilon^{ab}_{~~c}\null \! \mathop {\vphantom {N}\smash N}
\limits ^{}_{^\sim}\!\null \tilde{E}^j_{b} F_{ij}^{c}
+N^j F^a_{ji} +{\cal D}_i{\cal A}^a_{0}+e\Lambda \null \! 
\mathop {\vphantom {N}\smash N}\limits ^{}_{^\sim}\!\null e^a_i, 
\label{eqA} \\
\dot{\tilde{E}^i_a}
&=&-i{\cal D}_j( \epsilon^{cb}_{~~a} \null \! 
\mathop {\vphantom {N}\smash N}\limits ^{}_{^\sim}\!\null 
\tilde{E}^j_{c}
\tilde{E}^i_{b})
+2{\cal D}_j(N^{[j}\tilde{E}^{i]}_{a})
+i{\cal A}^b_{0} \epsilon_{ab}^{~~c} \tilde{E}^i_c,  \label{eqE}
\end{eqnarray}
\noindent
where 
${\cal D}_jT^{ji}_a:=\partial_jT^{ji}_a-i
 \epsilon_{ab}^{~~c} {\cal A}^b_{j}T^{ji}_c,$ 
 for $T^{ij}_a+T^{ji}_a=0$.

In order to construct metric variables from the variables 
$({\cal A}^a_i, \tilde{E}^i_a, \null \! 
\mathop {\vphantom {N}\smash N}\limits ^{}_{^\sim}\!\null, N^i)$, 
we first prepare 
tetrad $E^\mu_I$ as
$E^\mu_{0}=({1 / e \null \! \mathop {\vphantom {N}\smash N}
\limits ^{}_{^\sim}\!\null}, -{N^i / e \null \! 
\mathop {\vphantom {N}\smash N}\limits ^{}_{^\sim}\!\null})$ and
$E^\mu_{a}=(0, \tilde{E}^i_{a} /e).$
Using them, we obtain metric $g^{\mu\nu}$ such that
\begin{equation}
g^{\mu\nu}:=E^\mu_{I} E^\nu_{J} \eta^{IJ}. \label{recmet}
\end{equation}

\subsection{Reality conditions}\label{subsec:reality}

Notice that in general the metric (\ref{recmet}) is not real. 
In order to recover the real metric, 
we must impose the reality conditions.

To ensure the metric is real-valued,  
we need to impose two conditions;
the primary is that the doubly densitized contravariant metric
$\tilde{\tilde \gamma}{}^{ij} := e^2 \gamma^{ij}$ is real,
\begin{equation}
\Im (\tilde{E}^i_a \tilde{E}^{ja} ) = 0, \label{w-reality1} 
\end{equation}
and the secondary condition is that the time derivative of
$\tilde{\tilde \gamma}{}^{ij}$ is real,
\begin{equation}
\Im \{ \partial_t(\tilde{E}^i_a \tilde{E}^{ja} ) \} = 0.
 \label{w-reality2} 
\end{equation}

Using the equations of motion for $\tilde{E}^i_{a}$ (\ref{eqE}), 
the gauge constraint (\ref{c-g}) and 
the primary reality condition (\ref{w-reality1}), we can replace 
the secondary  condition (\ref{w-reality2})
with a different constraint 
\begin{equation}
W^{ij}:=\Re (\epsilon^{abc} 
\tilde{E}^k_a \tilde{E}^{(i}_b {\cal D}_k \tilde{E}^{j)}_c) 
\approx 0,  
\label{w-reality2-final}
\end{equation}
which fixes six components of ${\cal A}^a_{i}$ and $\tilde{E}^i_a$.
Moreover, in order to recover the original lapse function 
$N := \null \! \mathop {\vphantom {N}\smash N}
\limits ^{}_{^\sim}\!\null e$,
we demand $\Im (N/e)=0$, i.e. the density $e$ be real and
positive.
This requires that $e^2$ be positive, i.e.
\begin{eqnarray}
{\rm det}\tilde{E}>0. \label{t-reality1}
\end{eqnarray}
The secondary condition of (\ref{t-reality1}),
\begin{eqnarray}
\Im[\partial_t({\rm det}\tilde{E})]=0, \label{t-reality2}
\end{eqnarray}
\noindent
is automatically satisfied 
(see \cite{ys-con}).
Therefore, in order to 
ensure that $e$ is real, we only require (\ref{t-reality1}).

Rather stronger reality conditions are sometimes useful 
in  Ashtekar's formalism for recovering the 
real 3-metric and extrinsic curvature.  These conditions are 
\begin{eqnarray}
\Im (\tilde{E}^i_a ) &=& 0  
\label{s-reality1} \\
{\rm and~~} 
\Im  ( \dot{\tilde{E}^i_a} ) &=& 0, 
\label{s-reality2} 
\end{eqnarray}
\noindent
and we call them the ``primary triad reality condition" and the 
``secondary triad
reality condition", respectively.  
Using the equations of motion of $\tilde{E}^i_{a}$, 
the gauge constraint (\ref{c-g}),
the metric reality conditions (\ref{w-reality1}), (\ref{w-reality2})
and the primary condition (\ref{s-reality1}),
we see  that  (\ref{s-reality2}) is equivalent to \cite{ys-con}
\begin{equation}
\Re({\cal A}^a_{0})=
\partial_i( \null \! \mathop {\vphantom {N}\smash N}
\limits ^{}_{^\sim}\!\null )\tilde{E}^{ia}
+{1\over 2}e^{-1}e^b_i\null \! \mathop {\vphantom {N}\smash N}
\limits ^{}_{^\sim}\!\null\tilde{E}^{ja}\partial_j\tilde{E}^i_b
+N^{i}\Re({\cal A}^a_i). \label{s-reality2-final}
\end{equation}  From this expression we see that 
the second triad reality condition 
restricts the three components of ``triad lapse" vector 
${\cal A}^a_{0}$.
Therefore (\ref{s-reality2-final}) is 
not a restriction on the dynamical variables 
(${\cal A}^a_i$ and $\tilde{E}^i_a $)
but on the slicing, which we should impose on each hypersurface.
Thus the second triad reality condition does not restrict the 
dynamical variables any
further than the second metric condition does.


\ifx\answ\bigans
\newpage
\begin{figure}[h]
\setlength{\unitlength}{1in}
\begin{picture}(6.0,5.0)
\put(1.0,0.0){\epsfxsize=4.5in \epsffile{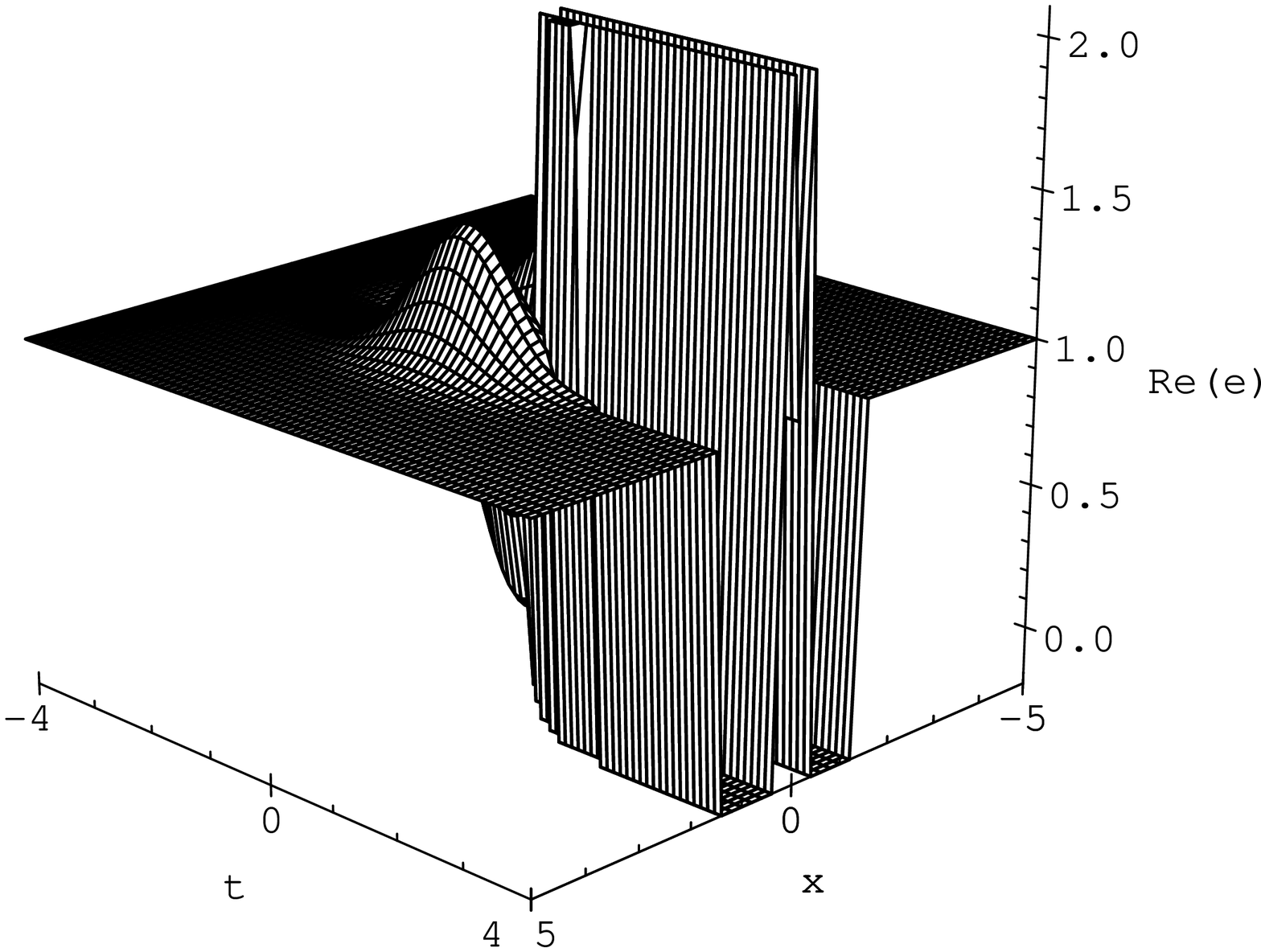} }
\end{picture}
\caption[ash-inter]{
An example of the {\it intersecting approach} for the dM metric 
(\ref{dM})
using Ashtekar's dynamical equations in evolutions. 
The density
is plotted versus spatial and time
comoving coordinates $x$ and $t$. 
A degenerate point exists at the origin
($t=x=0$), and we see that the time evolutions do not work properly
after intersecting the point.
}
\label{ash-inter}
\end{figure}

\begin{figure}[b]
\setlength{\unitlength}{1in}
\begin{picture}(6.0,8.0)
\put(1.0,4.0){\epsfxsize=4.5in \epsffile{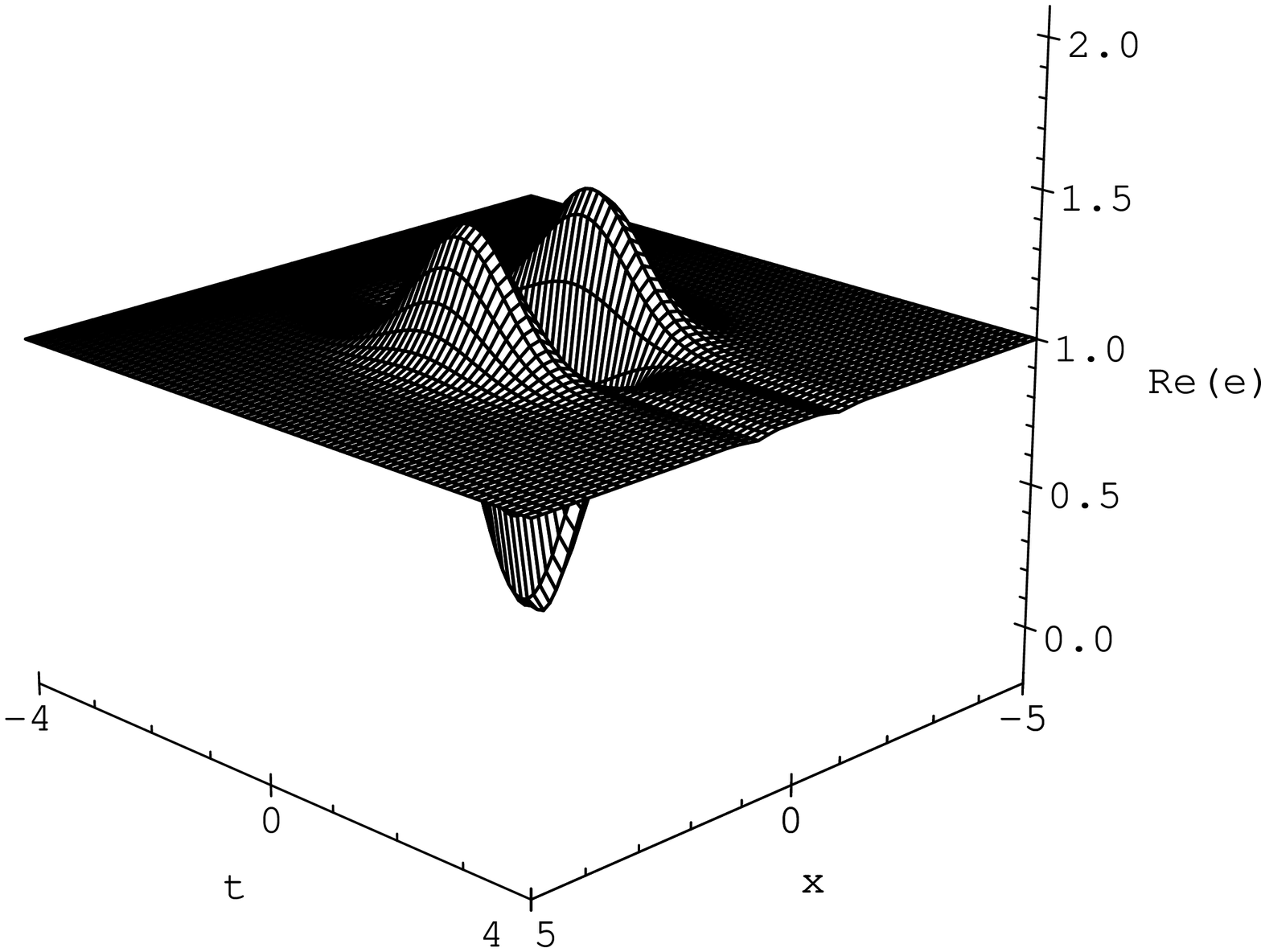} }
\put(1.0,0.0){\epsfxsize=4.5in \epsffile{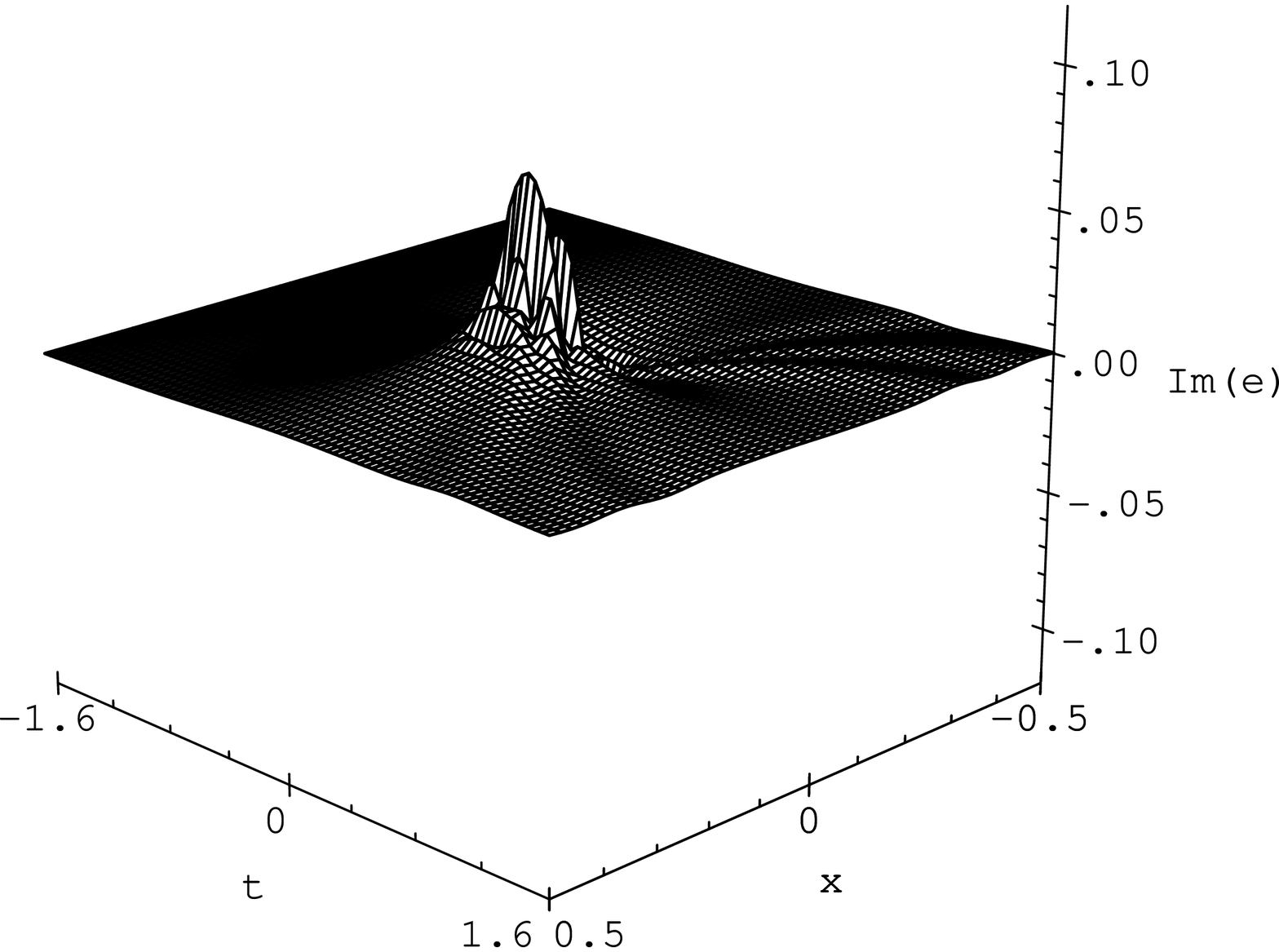} }
\end{picture}
\caption[ash-defo]{
An example of the {\it deformed slicing approach} for the 
dM metric (\ref{dM})
using Ashtekar's dynamical equation in evolution. 
The density
[real part(a) and imaginary part(b)] 
is plotted versus coordinates $x$ and $t$. 
A degenerate point exists at the origin
($t=x=0$), and we see that the time evolution works properly
for all the evolution and that the 3-metric satisfies
 the asymptotic reality condition again. 
}
\label{ash-defo}
\end{figure}
\fi

\ifx\answ\litans
\newpage
\begin{figure}[h]
\setlength{\unitlength}{1in}
\begin{picture}(4.0,4.0)
\put(1.0,0.0){\epsfxsize=3.0in \epsffile{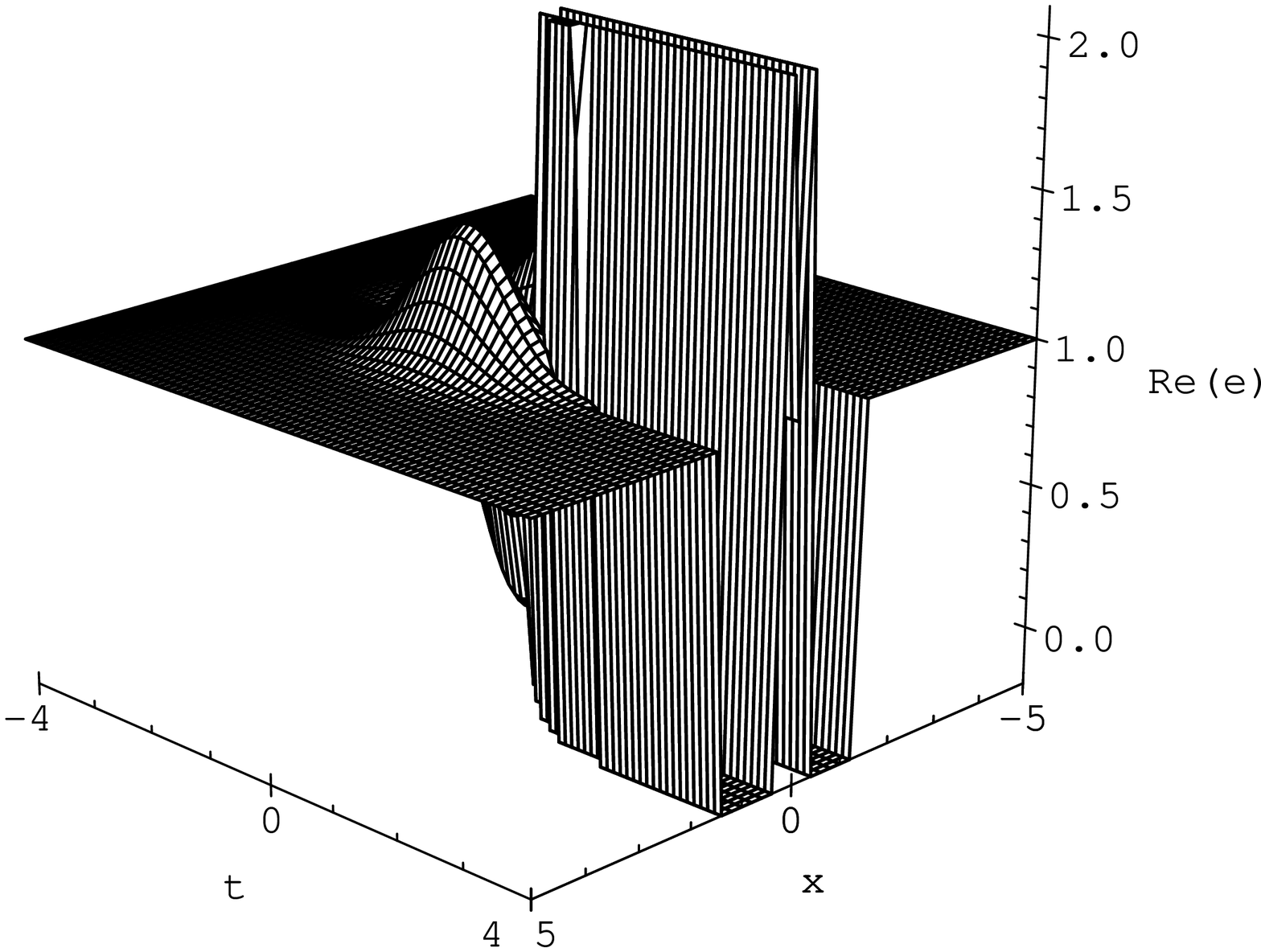} }
\end{picture}
\caption[ash-inter]{
An example of the {\it intersecting approach} for the dM metric 
(\ref{dM})
using Ashtekar's dynamical equations in evolutions. 
The density
is plotted versus spatial and time
comoving coordinates $x$ and $t$. 
A degenerate point exists at the origin
($t=x=0$), and we see that the time evolutions do not work properly
after intersecting the point.
}
\label{ash-inter}
\end{figure}

\newpage
\begin{figure}[b]
\setlength{\unitlength}{1in}
\begin{picture}(8.0,4.0)
\put(0.0,0.0){\epsfxsize=3.0in \epsffile{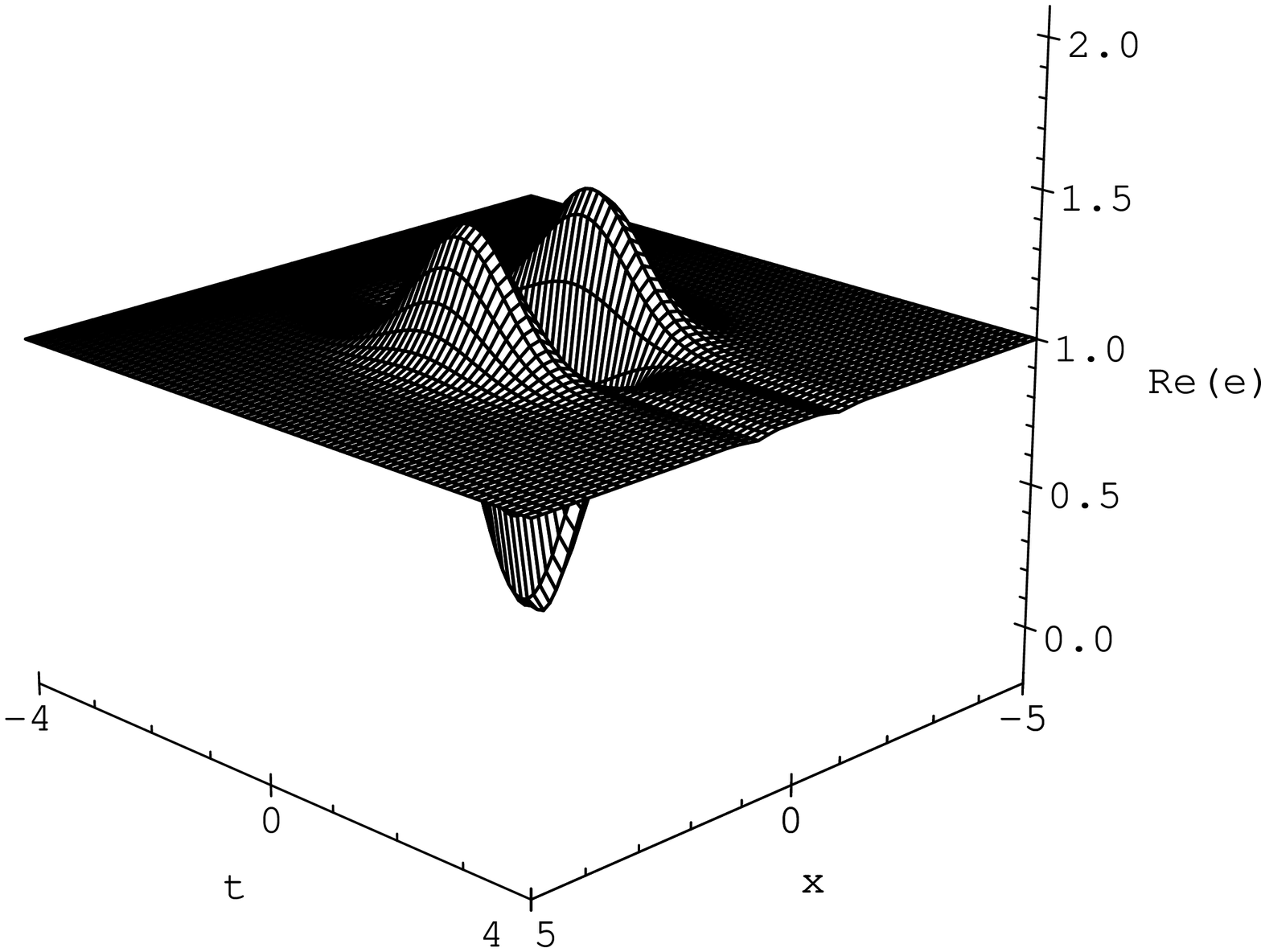} }
\put(4.0,0.0){\epsfxsize=3.0in \epsffile{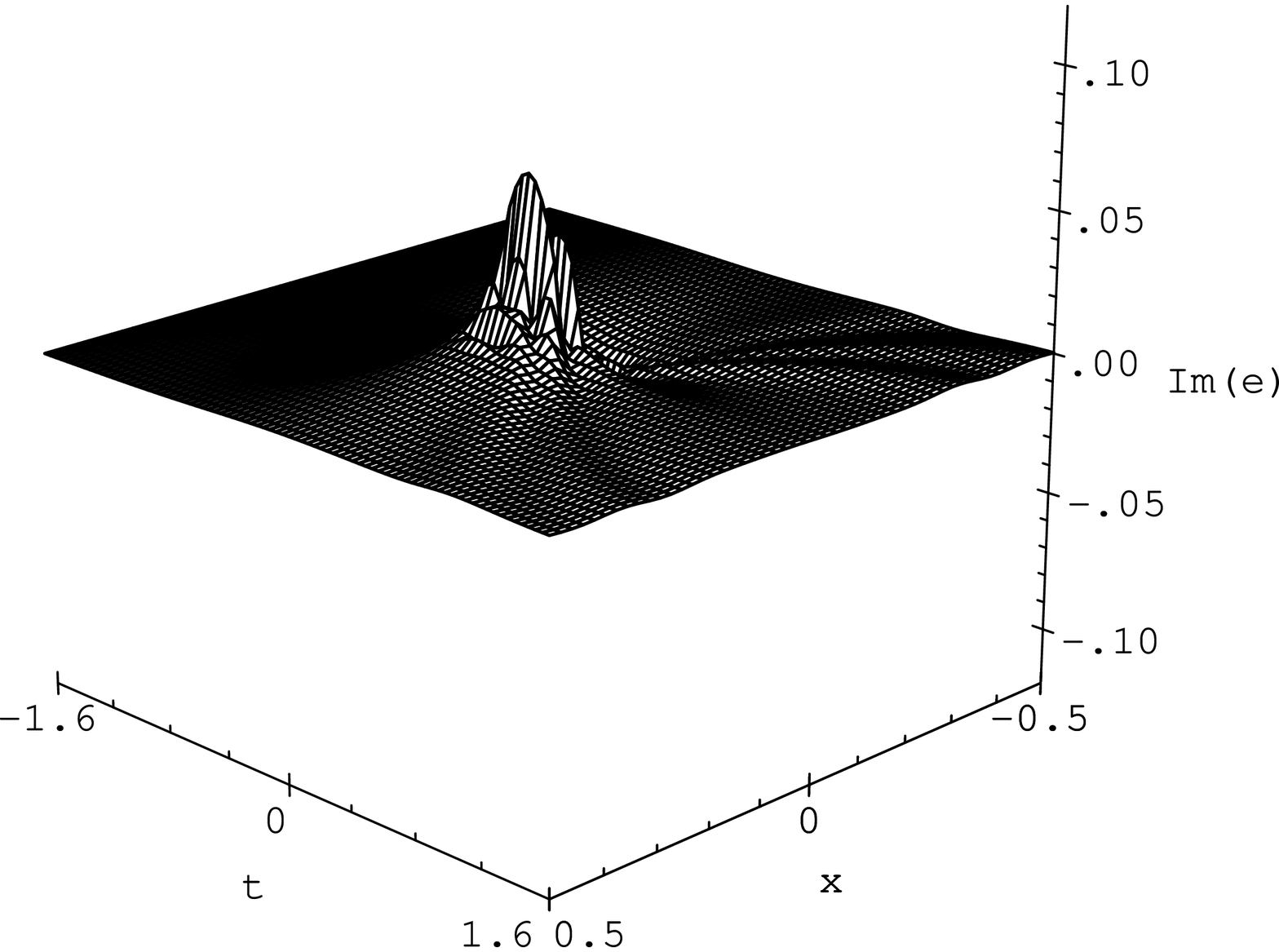} }
\end{picture}
\caption[ash-defo]{
An example of the {\it deformed slicing approach} for the 
dM metric (\ref{dM})
using Ashtekar's dynamical equation in evolution. 
The density
[real part(a) and imaginary part(b)] 
is plotted versus coordinates $x$ and $t$. 
A degenerate point exists at the origin
($t=x=0$), and we see that the time evolution works properly
for all the evolution and that the 3-metric satisfies
 the asymptotic reality condition again. 
}
\label{ash-defo}
\end{figure}
\fi

\end{document}